# A-STAR: The All-Sky Transient Astrophysics Reporter

J.P. Osborne[1], P. O'Brien[1], P. Evans[1], G.W. Fraser[1], A. Martindale[1], J.-L. Atteia[2,3], B. Cordier[4], S. Mereghetti[5]

**Abstract.** The small mission A-STAR (All-Sky Transient Astrophysics Reporter) aims to locate the X-ray counterparts to ALIGO and other gravitational wave detector sources, to study the poorly-understood low luminosity gamma-ray bursts, and to find a wide variety of transient high-energy source types, A-STAR will survey the entire available sky twice per 24 hours. The payload consists of a coded mask instrument, Owl, operating in the novel low energy band 4-150 keV, and a sensitive wide-field focussing soft X-ray instrument, Lobster, working over 0.15-5 keV. A-STAR will trigger on ~100 GRBs/yr, rapidly distributing their locations.

## 1. Introduction.

Responding to the 2012 call from the European Space Agency for a new type of small scientific mission for launch in late 2017, a consortium of institutes[6] proposed A-STAR: the All-Sky Transient Astrophysics Reporter. The mission call had a cost cap of €50M to ESA (including launch). Despite the severe cost limit, A-STAR is excitingly capable; it has three objectives:

1. Precisely locate the high-energy photon sources of gravitational-wave and neutrino transients and transients located by the new generation of astronomical facilities.
2. Reveal the physics underlying the variety in the population of gamma-ray bursts, including high-luminosity high-redshift bursts, low-luminosity bursts and short bursts.
3. Discover new high-energy transient sources over the whole sky, including supernova shock break-outs, black hole tidal disruption events, magnetar flares, and monitor known X-ray sources with daily observations.

## 2. Scientific Objectives.

2.1. Gravitational wave sources.

The launch of A-STAR will coincide with the true dawn of the era of multi-messenger astronomy during the second half of the current decade. The upgrading of the LIGO and VIRGO GW detectors (ALIGO/AVirgo) will revolutionize astronomy by permitting the detection and localization of GWs at a rate of perhaps dozens per year. IceCube and KM3NeT will likewise revolutionize neutrino astrophysics, routinely detecting cosmological neutrinos. Several of the most powerful sources of GWs predicted by general relativity, e.g. NS-NS or NS-BH mergers and core-collapse SNe, and neutrino sources such as GRBs produce powerful electromagnetic (EM) signals. ALIGO is planned to be operational by 2016-2018 and will be capable of identifying a randomly oriented NS-NS (NS-BH) merger out to ~450 (~900) Mpc, with a combined predicted rate of 50 yr$^{-1}$ (Abadie et al. 2010), but with relatively poor sky localizations of ~10-1000 sq. degrees (Klimenko et al. 2011). IceCube and KM3NeT can localize to an accuracy of better than a few degrees but within a smaller volume of the Universe (the IceCube Collaboration 2011; Kappes 2007). To maximize the science return of the multi-messenger era requires an in-orbit trigger and search facility that can either simultaneously detect the event or rapidly observe the large error boxes provided by the GW and neutrino facilities with good sensitivity to the EM signal. This combined requirement is uniquely provided by A-STAR, which is able to trigger using Owl or Lobster and observe a very large fraction of the GW/neutrino error boxes within an orbit due to the large grasp of the Lobster instrument compared to current generation X-ray facilities, e.g. Lobster has a grasp 40 times that of Swift/XRT. For events triggered on-board with Owl or Lobster, GW searches can also be carried on the resultant known sky locations with lower ALIGO/AVirgo signal-to-noise thresholds and hence an increased search distance.

[1] Dept of Physics & Astronomy, University of Leicester, LE1 7RH, UK
[2] UPS-OMP, Université de Toulouse, IRAP, Toulouse, France
[3] CNRS, IRAP, 14 avenue Edouard Belin, F-31400 Toulouse, France
[4] CEA, IRFU, Service d'Astrophysique, 91191 Gif-sur-Yvette, France
[5] INAF-IASF Milan, via Bassini 15, I-20133 Milano, Italy
[6] UK: Univ Leicester, ISIC; F: IRAP, LAM, ARTEMIS, CEA, IAP, APC; DK: DTU, SSC; I: IASF-Milan, IASF-Rome; B: Univ Liège; CH: ISDC; PL: SRC (PAS).

## 2.2. Gamma-ray bursts

In recent years GRBs have become essential in the study of stellar explosions, the evolution of massive stars (e.g. Pian et al. 2006), the extreme physics of relativistic jets and particle acceleration (e.g. Racusin et al. 2008), and as lighthouses illuminating the distant Universe (e.g. Tanvir et al. 2009). In the future we expect GRBs to provide a powerful probe of the epoch of reionization of the Universe (e.g. Tanvir et al. 2012), constrain the properties of the first stars, and will revolutionize GW astrophysics by associating GW signals with GRBs.

In this context, it is striking that so little is known about the origin of GRBs and the conditions needed for their production. Many basic questions remain unresolved: Why do only a few percent of Type Ibc supernovae produce long GRBs? What is the role of metallicity and binarity in LGRB production? Are some LGRBs powered by proto-neutron stars, at least initially? What is the range of beaming angles of long and short GRBs? What progenitor systems produce short GRBs? Why do some nearby long GRBs have no accompanying supernova (Fynbo et al. 2006; Gal-Yam et al. 2006)? Having a functioning orbiting high-energy facility to address these issues is essential, particularly when we consider the advent of future large observatories such as the European ELT and the JWST at visible and infrared wavelengths, the full ALMA interferometer in the sub-mm, SKA and its precursors in radio, and CTA and HAWC at very high energies, that will revolutionize our vision of GRBs and their host galaxies. Other time-domain surveys such as Pan-STARRS and LSST will bring the power of multi-wavelength observations to time domain astronomy, possibly revealing the elusive orphan afterglows of GRBs, permitting an accurate measure of GRB beaming.

## 2.3. Discovering new high-energy transients.

Magnetars, young NS with external magnetic fields of $10^{13}$–$10^{15}$ G, are among the most powerful and spectacular high-energy transients in the sky. About twenty sources believed to be magnetars are currently known in our Galaxy and in the Magellanic Clouds, but since most of them are transients with long quiescent periods, the total population waiting to be discovered is certainly much larger (Mereghetti 2008). Magnetars can produce Giant Flares thought to be due to star crustal fractures, we predict that A-STAR will see more than 2 per year. Intermediate flares are more frequent; the wide field of view and the frequent sky coverage of A-STAR will, for the first time, allow detection of a large number of flares and obtain a reliable estimate of the frequency of such events.

| Transient type | Rate |
|---|---|
| GW sources | 2-3 yr$^{-1}$ |
| GRB | 100 yr$^{-1}$ |
| Magnetars | 2 yr$^{-1}$ |
| SN shock breakout | 1 yr$^{-1}$ |
| TDE | 15 yr$^{-1}$ |
| AGN+Blazars | 100 day$^{-1}$ |
| Thermonuclear bursts | 10 day$^{-1}$ |
| Novae | 1 |
| Dwarf novae | 10 day$^{-1}$ |
| SFXTs | 10 yr$^{-1}$ |
| Stellar flares | ~100 yr$^{-1}$ |
| Stellar super flares | 1 week$^{-1}$ |

*Table 1. A-STAR detection rates for different astrophysical transients and variables*

The birth of a new SN is revealed by a burst of high-energy emission as the shock breaks out of the star. This has been spectacularly captured in a serendipitous Swift XRT observation of SN2008D (Soderberg et al. 2008): SNe are usually found only days to weeks after the explosion as radioactive heating powers optical brightening. Few observations exist early in a SN evolution: SN2008D remains the only non-GRB SN to be detected in X-rays at the time of first radiation escape from the star. A-STAR will significantly advance our understanding of the SN explosion mechanism, detecting SNe at the very moment of emergence, gathering comprehensive, prompt data and alerting follow-up communities to these landmark events.

Tidal disruption events (TDEs) offer a unique probe of the ubiquity of BH in galaxies, accretion on timescales open to direct study, and the nature and dynamics of galactic nuclei. Such events are expected to be visible as luminous, roughly Eddington limited objects with hot, UV and soft X-ray emission (e.g. Komossa et al. 2004; Gezari et al. 2012). The recent discovery by Swift of two highly luminous outbursts from galactic nuclei implies that at times a fraction of this energy is deposited in a new relativistic jet outflow (Levan et al. 2011; Bloom et al. 2011; Burrows et al. 2011), offering a new route to their identification and an opportunity to study newly-born jets. A-STAR is ideal for both the discovery and characterization of TDEs, opening new windows on numerous astrophysical questions.

## 3. A-STAR instruments.

### 3.1. Owl.

Owl is a coded mask telescope operating in the 4-150 keV energy range. It has a wide field of view (~1.44 sr) and a 10′ source error radius (90% confidence) for the faintest sources (7σ detection), improving to 2′ for the brightest (>30σ).

The Owl detector plane is made of 3840 Schottky CdTe detectors (4×4 mm × 1 mm thick) yielding a geometrical area of 614 cm$^2$. The new generation ASICs developed at CEA Saclay, together with the careful detector selection and the optimized hybridization done at IRAP Toulouse allow to lower the detection threshold with respect to former CdTe detectors by about 10 keV, reaching ~4 keV.
A coded aperture mask, made of a 0.6 mm thick tantalum sheet, placed 46 cm above this detection plane, defines a coded field of view of ≈ 60°×88°. In order to optimize the sensitivity for short bursts the mask aperture was set at 40%.

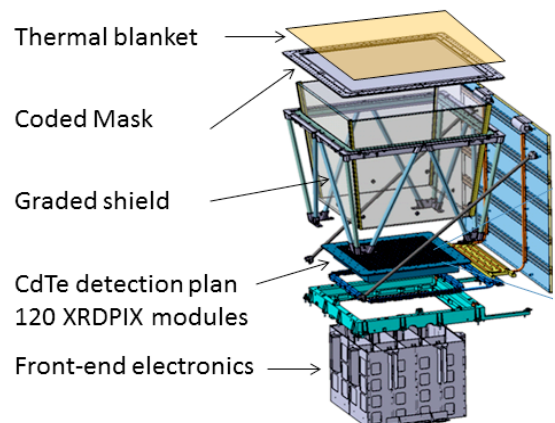

**Fig. 1.** The Owl coded mask instrument

Taking into account the geometrical parameters, and the materials present along the optical axis, the effective area is estimated to 104 cm$^2$ at 4 keV.

Data are continuously analyzed on board in order to detect bursts, by first detecting a count rate increase (in several energy bands) on time scales from 10 ms to 20 s followed by the formation of the image in the triggered time window in which a new source is searched, or by systematic searches for new sources in images built on time scales from 20 to 1200 s. Triggers are sent to the A-STAR service module to request a slew maneuver

By decreasing its low energy threshold in the soft X-ray domain, while keeping a large field-of-view, Owl will open a new window on the Universe. It will detect almost all of the GRB population seen by Swift, and is especially sensitive to highly red-shifted bursts and to the poorly understood low energy bursts.

## 3.2. Lobster.

Lobsters and other crustaceans focus by grazing incidence reflection off curved square pore optic arrays. This technique uniquely provides X-ray focussing over a very wide field, and is ideally suited to the A-STAR goals as it enables efficient detection of a large number of GRBs in a new low energy regime.

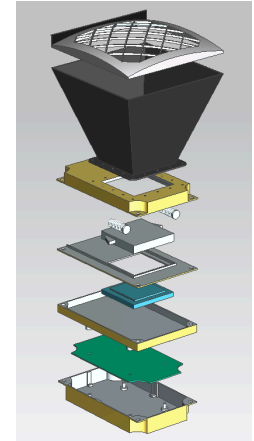

Lobster comprises 3 modules, each with a 17.3×17.3° field of view. These form a single 900 square degree FOV, centred on that of Owl. Each Lobster module has an array of 7×7 40×40 mm Micro-Channel Plate optics, mounted onto a titanium front end, supported by a carbon fibre mechanical structure. The optics have a spherical radius of 600 mm, focusing onto the detector at 300 mm. The camera contains an MCP detector which is curved to match the 300mm curvature of the focal plane, an anti-coincidence system, a thin aluminized polyimide optical/UV blocking filter, and the readout and analogue electronics. This configuration has a position resolution FWHM of 25 μm over the ~93 × 93 mm imaging area of the detector, sufficient to significantly oversample the PSF.

**Fig. 2.** A single Lobster module

Lobster provides un-vignetted and uniform resolution imaging across a very wide FOV, while maintaining the imaging advantage in sensitivity over collimating and coded mask systems. The point spread function of a Lobster optic is cruciform, 25% of the counts fall in the central peak, 50% in the arms and 25% in a diffuse centred pyramid. The detector performance is well matched to the optic capabilities and allows source position centroiding to <1.8' for 90% of GRBs, <0.5' for 50% and 10% better than <0.17' for 10% (determined by Monte-Carlo simulation of Lobster observation of Swift GRBs).

The position and trigger algorithm maximises sensitivity by employing two stages: the first projects the image in two perpendicular 1D histograms, and a candidate detection is when a 2.5σ event is seen above background in both axes. The second stage takes a cross shaped patch centred on the candidate and integrates over time, the transient is confirmed if the signal exceeds a specifiable higher statistical significance (e.g. ~6.3σ to achieve a false trigger rate of 1 in $10^{10}$). At this level the 30-minute sensitivity is $3\times10^{-11}$ erg.cm$^{-2}$.s (0.15-5 keV). The two-stage approach maximises sensitivity to faint transients and minimises false alerts.

## 4. Mission Profile.

A-STAR carries two wide-field X-ray imagers and a fast communication system. The payload is designed to operate with the mass and power resources provided by a microsatellite platform. We have performed detailed accommodation studies, which have demonstrated that the Myriade Evolutions and Proba satellites fit within a standard Vega piggyback volume. The operational mission life is 3 years.

The scientific return of A-STAR depends crucially upon the ability of the satellite to point the instruments at the open sky for 15−30 min long exposures; the ability to compute the positions of detected transients on-board and to transmit alerts quickly to the ground; and a fast reaction on reception of Targets of Opportunity. Key to success is the number of transients detected; the mission profile is optimized to maximize this number. We have therefore chosen an orbit with an altitude of 650 km and an inclination i<30° in order to minimize the time spent unusably in a high radiation environment. This orbit can be reached with a Vega launch as a passenger (A-STAR fits within the standard piggyback volume).

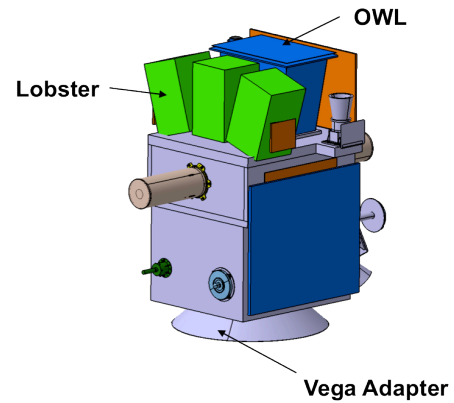

**Fig. 3.** A-STAR on Myriade Evolutions

The survey strategy is an essential ingredient of the mission. A-STAR will cover a significant fraction of the sky twice a day, perform long exposures taking full advantage of Lobster sensitivity, and observe Owl-detected transients with Lobster within 1-2 minutes. The A-STAR strategy relies on three ~1500 sec long dwells per orbit, two in the day side and one in eclipse, respecting the 90° sun avoidance angle required by Owl and Lobster. The eclipse dwells will detect transients in the night hemisphere, allowing a prompt response by ground-based facilities.

The need to transmit alerts quickly (within 1 minute) from the spacecraft to the ground calls for a dedicated system. We have studied two possible systems: in the first option, the consortium provides a network of VHF ground stations, as studied by CNES for the SVOM mission. An alternative is to use the COM Dev (Europe) SB-Sat system to send short messages via Inmarsat.

## 5. Concluding Remarks.

ESA announced that A-STAR was not selected on October 15, 2012. We believe that the science case for a wide-field high-cadence X-ray survey is very strong, and in particular that the promise of the new generation of gravitational wave detectors is best realised by such a survey. We intend to pursue future new mission opportunities vigorously.

**Acknowledgements.** We thank the 29 Co-Is and 62 associate scientists for their valuable input to the A-STAR proposal. JPO & PE acknowledge the support of the UK Space Agency.

**References**
Abadie J. Abbot, B.P., Abbot, R., et al., 2010, Class. Quantum Grav., 27, 173001
Klimenko S., Vedovato, G, Drago, M., et al., 2011, PhRvD, 83, 102001
The IceCube Collaboration, 2011, arXiv:1111.2741
Kappes A., 2007, arXiv:0711.0563
Pian E., Mazalli, P.A., Masetti, N., et al., 2006, Nature, 442, 1011
Racusin J.L., Karpov, S.V., Sokolowski, M., et al., 2008, Nature, 455, 183
Tanvir N.R., Fox, D.B., Levan, A.J., et al., 2009, Nature, 461, 1254
Tanvir N.R., Levan, A.J., Fructer, A.S., et al., 2012, ApJ, 754, 46
Fynbo J.P.U., Watson, D., Thöne, C.C, et al., 2006, Nature, 444, 1047
Gal-Yam A., Fox, D.B., Price, P.A., et al., 2006, Nature, 444, 1053
Mereghetti S., 2008, A&ARv, 15, 225
Soderberg A.M., Berger, E., Page, K.L., et al., 2008, Nature, 453, 469
Komossa S., Halpern, J., Schartel, N., et al., 2004, ApJ, 603, L17
Gezari S., Chornock, R, Rest, A., et al., 2012, Nature, 485, 217
Levan A.J., Tanvir, N.R., Cenko, S.B., et al., 2011, Science, 333, 199
Bloom J.S., Giannios, D., Metzger, B., et al., 2011, Science, 333, 203
Burrows D.N., Kennea, J.K., Ghisellini, G., et al., 2011, Nature, 476, 421